\documentclass[preprint]{vgtc}             

\usepackage{xurl}
\usepackage{tabu}                      
\usepackage{booktabs}                  
\usepackage{lipsum}                    
\usepackage{mwe}                       
\usepackage{mathptmx}                  
\usepackage{url}
\usepackage{xspace}
\usepackage{subcaption}
\usepackage{graphicx} 
\usepackage{multirow}
\usepackage{makecell}
\usepackage{xcolor}
\usepackage{colortbl}
\usepackage{amsmath}
\usepackage[normalem]{ulem}
\usepackage{enumitem}
\usepackage{wrapfig}
\usepackage{rotating}
\usepackage[colorlinks=true, linkcolor=blue]{hyperref}
\usepackage{booktabs}       
\usepackage{pdflscape}
\usepackage[linesnumbered,ruled,vlined, noend]{algorithm2e}

\vgtccategory{Research}

\preprinttext{To appear in the IEEE Conference on Virtual Reality and 3D User Interfaces (IEEE VR) 2026.}

\title{What Sensors See, What People Feel: An Exploratory Study of Subjective Collaboration Perception in Mixed Reality}

\author{
Yasra Chandio\thanks{Yasra Chandio and Diana Romero contributed as co-first authors.}\\
ychandio@umass.edu \\
University of Massachusetts Amherst\\
\and
Diana Romero\textsuperscript{*} \\
dromero@uci.edu \\
University of California, Irvine\\
\and
Salma Elmalaki \\
salma.elmalaki@uci.edu \\
University of California, Irvine\\
\and
Fatima M. Anwar \\
fanwar@umass.edu \\
University of Massachusetts Amherst}

\onlineid{1296}
\vspace{-.2cm}
\teaser{
  \centering
\includegraphics[width=0.6\textwidth]{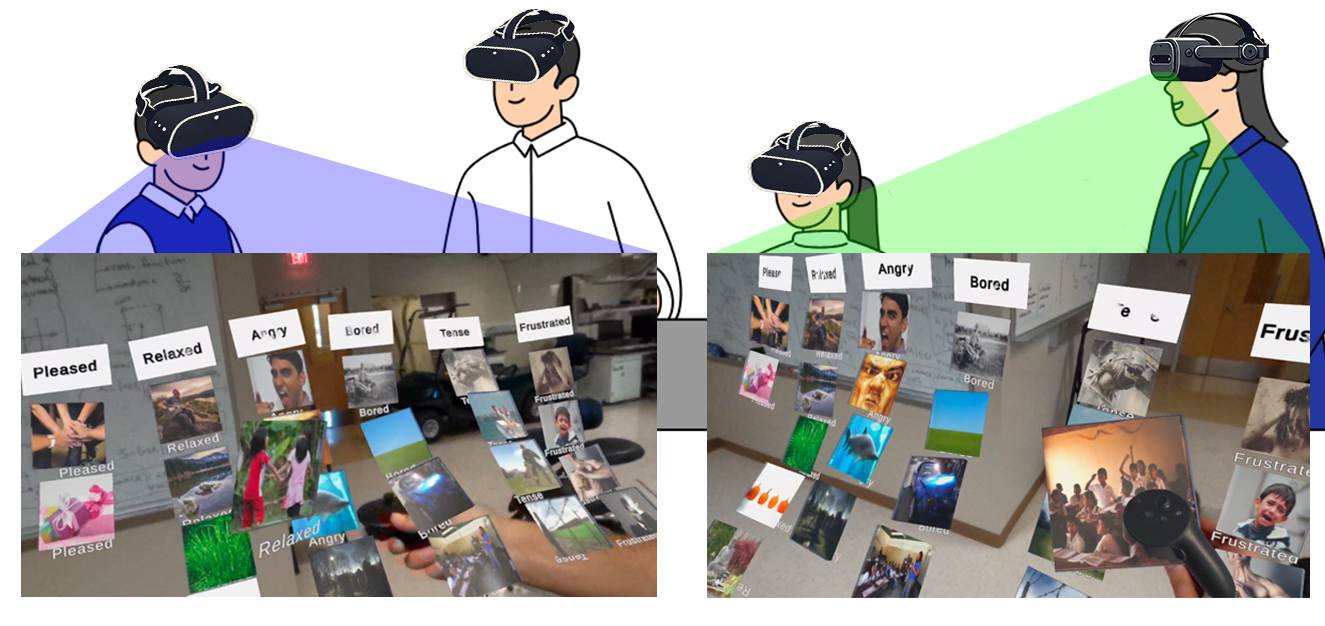}
\vspace{-.2cm}
~
\includegraphics[width=0.28\textwidth]{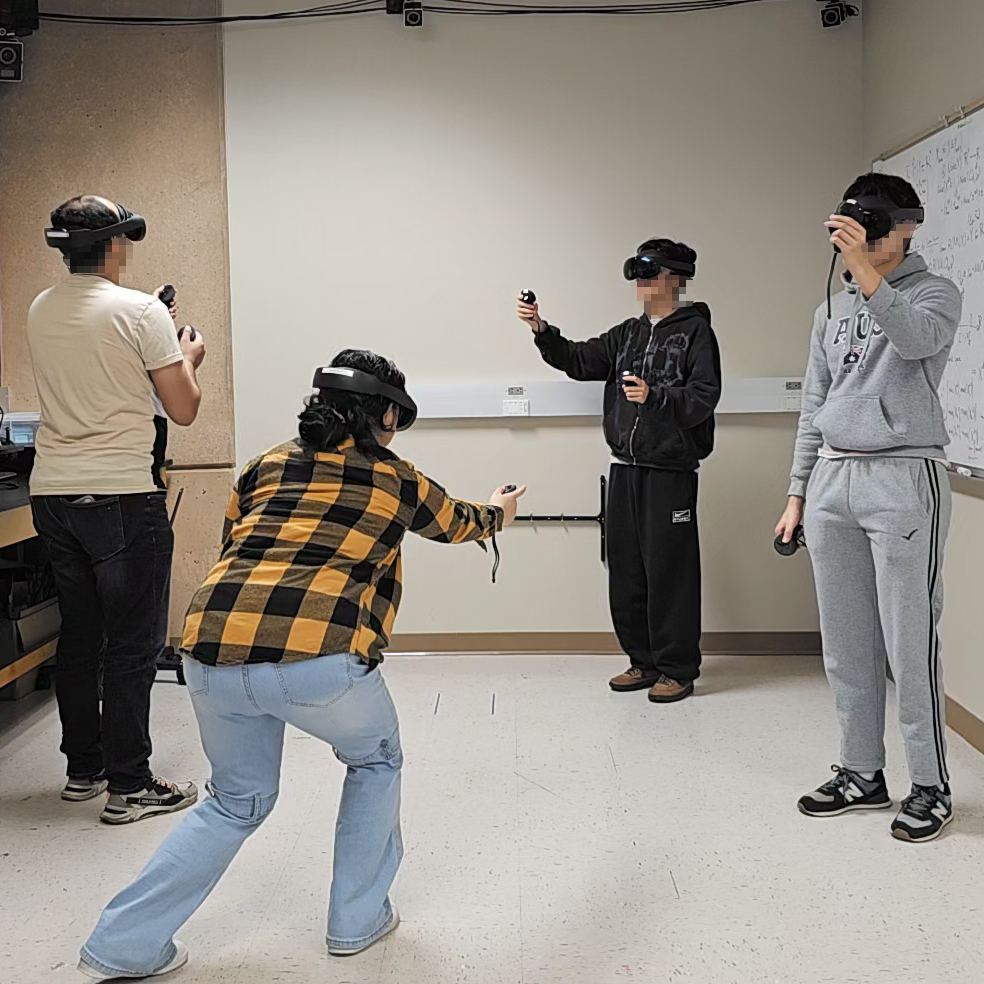}
\vspace{-.15cm}
  \caption{Illustration of the Collaborative MR Image Sorting Application. Left and Middle: First-person perspective of the application interface as experienced by two participants. Right: Third-person view showing participants interacting with the application in a shared MR environment.}
\vspace{-.3cm}
  \label{fig:teaser}
}

\abstract{
  Mixed Reality (MR) enables rich, embodied collaboration; however, it is uncertain whether sensor- and system-logged behavioral signals capture how users experience that collaboration. This disconnect stems from a fundamental gap. Behavioral signals are observable and continuous, while collaboration is interpreted subjectively and shaped by internal states like presence, cognitive availability, and social awareness. Our core insight is that sensor signals serve as observable manifestations of subjective experiences in MR collaboration, and they can be captured through sensor data such as shared gaze, speech, spatial movement, and other system-logged performance metrics. 
  We propose the Sensor-to-Subjective (\texttt{S2S}) Mapping Framework, a conceptual model that links observable interaction patterns to users' subjective perceptions of collaboration and internal cognitive states through sensor-based indicators and task performance metrics. To evaluate this model, we conducted an exploratory study with 48 participants across 12 MR groups engaged in a collaborative image-sorting task. 
  Our findings show a correlation between sensed behavior and perceived collaboration, particularly through shared attention and proximity.  
} 

\nocopyrightspace

\begin{document}
\vspace{-2mm}
\firstsection{Introduction}
\label{sec:introduction}
\maketitle
Collaboration in mixed reality (MR)\footnote{\textbf{MR} aligns with Milgram's reality-virtuality continuum~\cite{milgram1995augmented}, closely related to Augmented Reality (AR), where virtual elements seamlessly integrate with the real world.} 
 environments are becoming increasingly prevalent across domains such as education, design, healthcare, and remote work~\cite{ensRevisitingCollaborationMixed2019}. MR uniquely blends physical and virtual realms, allowing co-located and remote users to share digital content seamlessly. As MR becomes more immersive and integral to collaboration, researchers have investigated the effects of gaze, gesture sharing, and virtual replicas on the presence and cognitive load~\cite{bai2020user-eye-hand-gesture-collaboration, 10049700}.
Despite these advances, a fundamental challenge persists: \emph{How can we effectively understand the quality of collaboration in MR?}

Modern MR systems generate extensive behavioral data through system logs and embedded sensors, capturing where users look~\cite{10765374}, how they move~\cite{10.1145/3385956.3418968}, and when they speak~\cite{10.1145/3447993.3483272}; however, these signals often provide only limited insights into users' internal states~\cite{zhang_teamsense_2018}. Prior work shows that speech, gaze, and turn-taking are tightly connected to the internal perception of collaboration~\cite{microsoft-collaboration}. However, two core limitations remain: (1) it is unclear whether sensor data meaningfully reflects how participants perceive collaboration, or if key experiential dimensions go undetected. (2) Existing studies demonstrate specific bi-modal correlation mappings, such as gaze to presence~\cite{bai2020user-eye-hand-gesture-collaboration}, and physiological synchrony to co-presence~\cite{bayro2022subjective-objective-collaboration}, but stop short of offering a holistic objective-to-subjective mapping of collaboration. 
Bridging this gap is necessary to move from passive activity logging to human-aware MR systems that can infer and support collaboration quality in real time without relying on intrusive self-reports.

We address these gaps through a systematic, multimodal Sensor-to-Subjective (\texttt{S2S}) mapping framework, applied in a multi-perspective study of collaborative MR involving 48 participants across 12 groups.
First, \emph{we analyze sensor-derived indicators of group interaction}, including gaze-based shared attention, conversation dynamics through audio, and spatial proximity, which can be captured through embedded sensing in MR headsets~\cite{zhang_teamsense_2018, yangImmersiveCollaborativeSensemaking2022}. 
Second, \emph{we examine task performance using system logs}, measuring timing, interaction patterns, and decision changes throughout the collaboration. Finally, \emph{we assess subjective experience through post-exposure questionnaires} that capture participants' perceptions of group dynamics as well as their individual sense of presence and cognitive effort. 
Together, these perspectives allow us to ask: \emph{\textbf{Do observable group interaction patterns, as captured by MR headsets, reflect how participants experience collaboration?}} As an exploratory study, our objective is to identify promising alignments and generate hypotheses for future confirmatory research, rather than establishing a final predictive model for all MR contexts. We formalize this inquiry through three research questions:
\begin{enumerate}[leftmargin=0.93cm, itemsep=-0.1cm, topsep=-0.1cm]
    \item[\textbf{RQ1:}] 
    Do sensor-based indicators align with perceived subjective collaboration in MR? 
    \item[\textbf{RQ2:}] How do sensor-based behavior indicators relate to task performance in a collaborative MR task?
    \item[\textbf{RQ3:}] What do objective sensing and individual subjective experience reveal about group behavior in MR?
\end{enumerate}

\vspace{-0.1cm}
\section{Background and Related Work}
\label{sec:background}
\subsection{Group Behavior in Immersive Environments}
\label{sec:rw-group-behavior}

Understanding collaborative group behavior has evolved from traditional settings to MR environments, where users interact in novel ways across co-located, remote, and hybrid configurations. \textbf{\emph{Group behavior}}, is defined as all forms of interaction and activity within a group~\cite{hackman2010group-main-paper}, and \textbf{\emph{collaboration}}, the intentional, coordinated effort among members to work together toward shared goals~\cite{hackman2010group-main-paper}, takes distinctive forms in MR contexts due to unique factors including free movement~\cite{podkosovaCopresenceProxemicsShared2018}, embodied gesture~\cite{jingImpactSharingGaze2022b}, co-presence~\cite{slater2000small}, and shared spatial context~\cite{piumsomboon2019effects}. Research has examined MR collaboration through interface design~\cite{bailenson2008use}, task coordination~\cite{yee2007proteus}, communication patterns~\cite{abdullahVideoconferenceEmbodiedVR2021}, and trust development~\cite{bailenson2004transformed}, but it generally relies on external observations or outcomes rather than internal experiences of the participants. This creates \emph{a critical knowledge gap on how participants' internal experiences align with the observable behaviors logged by MR systems}.

 \subsection{Sensor-Based Indicators of Group Behavior}
 \label{sec:rw-sensors}
Sensor-based methods across ubiquitous and immersive computing characterize group behavior through multiple data streams (gaze, voice, motion, location) without relying on external observation or self-reporting. Remote collaboration systems, such as TeamSense~\cite{zhang_teamsense_2018} and CoCo~\cite{samroseCoCoCollaborationCoach2018} monitor nonverbal cues to track group cohesion, while wearable systems assess synchrony and proximity patterns during collaborative tasks~\cite{sunUsingWearableSensors2023}. 
In MR, headset-embedded sensors can track gaze, object manipulation, and gesture alignment, which researchers have shown correspond to collaboration quality~\cite{yangImmersiveCollaborativeSensemaking2022,irlittiVolumetricMixedReality2023}. Studies have correlated gaze and gesture sharing with co-presence~\cite{bai2020user-eye-hand-gesture-collaboration}, and linked physiological synchrony and gaze with perceived collaboration quality~\cite{bayro2022subjective-objective-collaboration}. Other systems enhance spatial awareness by embedding gaze cues~\cite{eyeMR-Vis-2021}, while multimodal integrations of gaze and gesture have been used to improve performance and presence~\cite{bailensonGazeTaskPerformance2002}. Additional work shows high-accuracy cognitive load prediction using multimodal sensing~\cite{hou2025cognitive-multimodal-sensor} and demonstrates that even minimal AR cues can influence team experience~\cite {sensor-team-work}. 
Conversation-focused research further emphasizes how speech alignment, gesture, gaze, and turn-taking synchronize to create effective collaboration in co-located settings~\cite{thomas2023communication-collaboration, microsoft-collaboration}. These findings emphasize that \emph{collaboration is fundamentally multimodal}, involving a convergence of signals across time. 

Despite these insights, existing work remains limited in two ways: (1)  Bi-modal Isolationism: isolated pairwise mappings (such as gaze to presence or physiological synchrony to co-presence) rather than producing a unifying mapping model and treating these as independent variables, and (2) Outcome-Centricity: lens-like views often aim to enhance either performance outcomes or structural interaction properties (such as coordination patterns and role distribution) without examining their relationship to participants' subjective experiences. \emph{Our approach differs not simply by aggregating these modalities, but by shifting the focus from pairwise correlations to triadic alignment}. 

Prior work has explicitly recognized the need for structured frameworks that integrate multiple sensing modalities and interpret internal collaboration states. Barrett et al. call for multi-method frameworks that integrate sensor modalities to capture internal collaboration states~\cite{rosen2015integrative-sensor-teamwork-healthcare}, while social signal processing emphasizes the need to fuse multimodal behavioral signals to capture fine-grained group dynamics~\cite{sensor-team-work}. Systematic reviews also note the absence of frameworks that link objective data to users' subjective experiences~\cite{de2019systematic-collabMR-survey}. Despite these advances, the critical question of whether sensor-based behavioral indicators meaningfully reflect users' internal perceptions of collaboration remains largely unexplored, particularly in co-located MR tasks using commercial headsets. Unlike a collection of bi-modal studies that treat sensors as stand-alone predictors, in this work, we position sensors, performance, and perception as interdependent nodes that validate one another through a unified interpretive lens.

\vspace{-0.1cm}
\subsection{Subjective Indicators of Group Behavior}
\label{sec:rw-subjective-measures}
Subjective self-report surveys remain central to understanding collaborative experiences. Standardized tools, such as post-task surveys, presence questionnaires~\cite{witmer_measuring_1998, schubert_experience_2001}, and cognitive load surveys~\cite{hart_nasa-task_2006}, and team cohesion surveys \cite{zhang_teamsense_2018} gauge how users internalize virtual social presence \cite{poeschl2015measuring-socail-copresence}, mutual understanding \cite{yiqiuzhouCharacterizingJointAttention2022}, and contribution equity \cite{voletIndividualContributionsStudentled2017}, while collaborative design work leverages them to assess idea sharing and creative synergy~\cite{paulus2000groups}. Despite these rich insights, subjective measures are rarely linked systematically to sensor-based behavioral data. We address this disconnect by investigating whether MR-embedded signals (gaze, speech, proximity) meaningfully reflect these internal states, positioning this alignment as a core concern for human-centered sensing.

\emph{To this end, in this paper, we study how sensor-based indicators of group behavior relate to subjective perceptions in collaborative MR. Instead of treating system-logged signals as stand-alone metrics, we treat them as potential reflections of collaborative experience, positioning this alignment as a core concern for human-centered sensing in MR.}

\vspace{-0.1cm}
\begin{figure}
    \centering
    \includegraphics[width=0.8\linewidth]{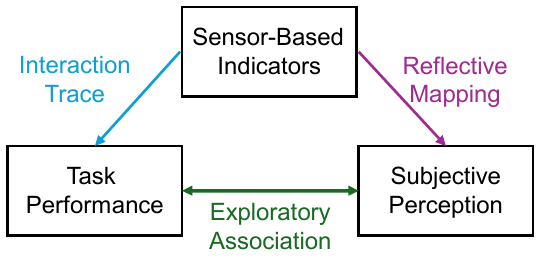}
    \vspace{-0.35cm}
    \caption{Sensor-to-Subjective Mapping Framework (S2S)}
    \label{fig:model}
\vspace{-0.4cm}
\end{figure}
\section{The Sensor-to-Subjective (S2S) Mapping Framework}
\label{sec:approach}

We propose the \texttt{Sensor-to-Subjective (S2S)} mapping framework to structure our investigation~(\autoref{fig:model}). \texttt{S2S} is an \emph{interpretive alignment model} and theoretical structure designed to \uline{explain} how high-frequency behavioral data relate to users’ internal collaborative experiences. The primary motivation for this alignment is the transition from passive logging to \uline{exploring} active intervention.
Rather than predicting subjective states from sensor data, we test whether and where system-observable sensor-derived interaction traces serve as an interpretive anchor for how collaboration is experienced. 
With \texttt{S2S}, we conceptualize collaborative group behavior across three interconnected perspectives (nodes): user-reported post-task reflections (subjective perceptions), observable system-level interaction traces (sensor-based indicators), and emergent outcomes from interaction (task performance metrics). We provide a foundation for MR systems to monitor, evaluate, and eventually enhance group dynamics autonomously. The critical distinction between \texttt{S2S} and existing bi-modal approaches that assume one view determines another~\cite{bai2020user-eye-hand-gesture-collaboration, bayro2022subjective-objective-collaboration, rosen2015integrative-sensor-teamwork-healthcare, thomas2023communication-collaboration}, is the shift from linear prediction to triadic synchronization. While prior work might ask 'Does gaze predict performance?', we ask \emph{'Does the sensor-performance alignment match the user's subjective reality?'}.

\subsection{Framework Nodes: Indicators and Measures}
\label{sec:theory-model}
\vspace{-0.15cm}
\subsubsection{Subjective Node: Perceptions of Collaboration}
\label{sec:subjective-collab-perception}
We define collaboration as not just joint actions, but how individuals interpret, feel, and reflect on their shared experience. In MR, social cues, task progress, and attention are distributed across a blend of physical and virtual elements, shaped by both individual states and group-level awareness. We treat this as the \emph{subjective perception of collaboration} and operationalize it through \emph{individual-level} and \emph{group-level} dimensions.

\vspace{0.09cm}
\noindent \textbf{Individual-Level Experience.} Each user’s interpretation of collaboration is filtered through their \emph{sense of presence}~\cite{perfromance-presence}, \emph{cognitive availability}~\cite{cognitiveload-presence}, and \emph{personal contribution}~\cite{chandio-vr-24-human-factors}. These passive background states directly influence how users perceive and make sense of social interaction~\cite{poeschl2015measuring-socail-copresence, yaremych2019tracing, williamson2021proxemics}. At its most foundational, collaboration begins with how individuals experience being part of a group in each moment. In immersive settings, this includes: 
\begin{itemize}[leftmargin=0.25cm, itemsep=-0.1cm, topsep=-0.1cm]   
\item \emph{Presence:} A user’s sense of immersion (``being there'') shapes their sensitivity to others’ gaze, gesture, and speech~\cite{slater2009place, schubert_experience_2001}. In our context, higher presence enhances awareness of others' engagement and informs group-level reflection.
\item \emph{Cognitive Load:} The mental effort to manage task demands affects users’ ability to perceive coordination signals. A taxed user may miss coordination cues, while a cognitively available one may detect attention shifts or conversational breaks~\cite{chandio2025Distraction}. Thus, we hypothesize that cognitive load influences not only task performance~\cite{kocur2020effects-external-perception-cognitive-performance} but also perceived collaboration quality~\cite{tang2022experiencing-extraneous-load}.
\item \emph{Contribution Awareness:} Users’ sense of how meaningfully others contributed reflects perceived participation balance~\cite{carroll2006awareness-teamwork}. This acts as a bridge between personal effort and social perception. It links internal judgment with observed peer input and becomes especially salient in unstructured, role-free MR settings.

\end{itemize}
\vspace{0.15cm}
These experiences are not isolated metrics but are shaped by real-time interpretations of others’ behaviors, and together they form the perceptual and cognitive basis for group-level reflections.

\vspace{0.15cm}
\noindent \textbf{Group-Level Reflection.} MR system tracks users individually, while users form post-task judgments about shared experience~\cite{steedLeadershipCollaborationShared1999}, such as whether they attended to the same objects, communicated effectively, and collaborated cohesively. We capture this through four constructs: 
\begin{itemize} [leftmargin=0.4cm, itemsep=-0.1cm, topsep=-0.07cm]
    \item \emph{Shared Attention Awareness} captures whether users felt others focused on the same virtual elements, indicating shared intent and coordination~\cite{mooreJointAttentionIts1995}.
    \item \emph{Conversational Support} reflects whether dialogue helped users understand (clarify) and contribute (invite participation), signaling cohesion and mutual support~\cite{goodwin1990conversation-analysis}.
    \item \emph{Proximity Impact} assesses how physical closeness, via MR, affected collaboration responsiveness. Prior work suggests that co-located interaction, being physically near others, improves responsiveness~\cite{gonzalez2015framework-proximity}.
    \item \emph{Group Collaboration} measures users' overall sense of group effectiveness based on observed behaviors, internal judgments, and implicit comparison to collaborative expectations~\cite{forsyth2021-group-cohesion}.
\end{itemize}

\vspace{0.15cm}
These individually collected measures reflect each user’s interpretation of the group, shaped by their state, engagement, and task involvement, forming \textbf{subjective readings of a social process.}

\subsubsection{Objective Node: Sensor-Based Indicators}
\label{sec:objective-collab-perception}
While subjective reports capture how collaboration feels, immersive systems log how it unfolds. Although these are individual-level measurements, they are not isolated; they reflect \emph{what each user does in relation to others in the group}. The same user performing the task alone would not produce the same sensor traces~\cite{hou2025cognitive-multimodal-sensor}. These sensor-based signals (derived from sensor data and system instrumentation) are shaped by interaction: who a user looks at, how close they are to others, and when they speak or remain silent. Interpreting such distributed traces in group contexts is key to understanding collaboration objectively. We focus on three core indicators supported by prior work~\cite{yangImmersiveCollaborativeSensemaking2022, zhang_teamsense_2018} and accessible on commodity MR headsets~\cite{MetaQuestProa, hololens2, AppleVisionPro}:
\begin{itemize}[leftmargin=0.35cm, itemsep=-0.1cm, topsep=-0.07cm]
\item \emph{\textbf{Conversation}} and speech activity have long been used to study group behavior across psychology, anthropology, and HCI~\cite{levine1990progress, brown2019group}. Recent work using ubiquitous sensors underscores the continued importance of conversation~\cite{leeFlowerPopFacilitatingCasual2018}. In MR, where interaction blends physical and virtual contexts, speech activity offers valuable insight into collaboration~\cite{yangImmersiveCollaborativeSensemaking2022}. To this end, we extract speaking time, turn count, and participation variance (non-transcribing content) via the headset microphone. These metrics, known to reflect dominance and fluency~\cite{pentland2012new}, offer insight into how evenly members contribute with privacy preservation.
 \item \emph{\textbf{Shared attention}}, the ability to jointly focus on an object underpins coordination and social connection~\cite{wolfJointAttentionShared2016}. It has been widely studied in collaborative systems~\cite{bakemanCoordinatingAttentionPeople1984, mooreJointAttentionIts1995}, often through gaze-based detection\cite{yiqiuzhouCharacterizingJointAttention2022}. We build on this by using MR headset eye-tracking to detect overlapping gaze on the same virtual object, enabling continuous, time-resolved logging of shared attention~\cite{mooreJointAttentionIts1995}. Such alignment reflects group awareness and may influence perceptions of collaboration, even without explicit communication. We aim to assess whether sensed shared attention reflects users' perception of collaboration. 
\item \emph{\textbf{Proximity}} influences collaboration by shaping knowledge exchange, creativity, and group cohesion~\cite{vissersKnowledgeProximity2013}. In co-located collaborative scenarios, groups tend to perform better when members are physically near one another rather than dispersed~\cite{hawkeyProximityFactorImpact2005, valacich1994physical}. Studies have observed that individuals often move closer to those with whom they share stronger social bonds~\cite{cristaniComputationalProxemicsInferring2011}. We capture proximity as the average pairwise distance between users using headset position data in a shared coordinate frame.
\end{itemize}

\vspace{0.15cm}

We focus on group-level distributions of proximity and conversation (overall closeness or speech participation) rather than directional relationships (who was closest to whom or who spoke to whom). Our goal is to capture collective engagement, not dominant roles. Prior work shows that proximity itself (regardless of direction) signals social engagement~\cite{cristaniComputationalProxemicsInferring2011}, and evenly distributed proximity and turn-taking indicate group cohesion~\cite{zhang_teamsense_2018}. We focus on subjective perceived group collaboration that is more likely influenced by general closeness than by specific spatial arrangements. 

\subsubsection{Performance Node: Emergent Task Outcomes}
\label{sec:performance-node} 
Beyond individual sensor signals and subjective reports, the \texttt{S2S} framework incorporates task-relevant interaction logs to measure the emergent outcomes of collaboration. These metrics serve as objective benchmarks for group efficiency and coordination complexity. We track three primary indicators (task details in \S\ref{sec:primary-task}): 
\begin{itemize}[leftmargin=0.35cm, itemsep=-0.1cm, topsep=-0.07cm] 
\item \emph{Interaction balance and complexity} log how often users interact with virtual content, the frequency of placement revisions (label changes), and the number of unique configurations explored before consensus. We focus on interaction balance rather than raw activity volume because participation equity reflects group cohesion and collective intelligence~\cite{woolley2010evidence}. High activity can come from a single dominant user, while balanced interaction indicates distributed coordination toward shared goals. High variance in these interactions can reflect participation inequity or negotiation difficulty.
\item \emph{Coordination fluency} monitors label overrides, which occur when one participant modifies a peer's previous decision. This serves as a proxy for group negotiation and the alignment of shared mental models. 
\item \emph{Temporal efficiency} captured by task completion time is recorded as a coarse but vital measure of overall collaborative success. 
\end{itemize} 
\vspace{0.1cm}
These metrics allow us to evaluate whether specific behavioral data (participation balance, coordination, and negotiation of decisions) lead to more coherent and faster group outcomes.

\subsection{Individual Reflections as Interpretive Anchors}
\label{sec:individual-reflections}
A central question in our exploratory study is how individual perceptions of collaboration relate to sensed behavioral traces. Rather than averaging subjective reports into a ``single group state'', we examine the distribution of perceptions across the group to identify alignment or divergence. Do some users feel disengaged or overloaded? These differences are not treated as noise; they reflect varied interpretations of what the collaboration felt like to the user, such as whether they felt disengaged or cognitively overloaded. By linking subjective internal perceptions (\S\ref{sec:subjective-collab-perception}) with sensor-based indicators (\S\ref{sec:objective-collab-perception}) and task performance metrics (\S\ref{sec:performance-node}), we ask: Do shared gaze events map onto users' awareness of shared attention? Does proximity reflect perceived cohesion? Do balanced interactions reflect users' perception of their own and peers' meaningful contributions to the task?

Importantly, in \texttt{S2S}, subjective states are not treated as downstream outcomes but as \textbf{interpretive anchors} that ground how sensor traces and task outcomes are understood; independent but complementary accounts of collaboration. This allows us to understand what groups did and how that behavior was experienced, yielding insights into how systems might better reflect, support, or adapt to group behavior in MR environments in the future. Together, these indicators form a continuous stream of behavioral data, interpreted as isolated features but also as \emph{observable patterns of interaction} that may or may not align with users' subjective interpretations of collaboration, avoiding the false precision that arises when sensors are interpreted in isolation.
Rather than labeling groups as ``effective'' or ``cohesive,'' our goal is to assess whether and how the system’s observations reflect how collaboration was experienced.

\subsection{Functional Relational Pathways}
\vspace{-0.095cm}
\label{sec:framework-pathways} 
The substance of the \texttt{S2S} framework lies in the functional mappings between its nodes, which allow us to explore the alignment between behavioral data and human experience. We define three primary pathways that guide the structure of our investigation:

\subsubsection{Reflective Interaction Mapping (Sensor $\leftrightarrow$ Subj.)}
We first examine whether sensor-derived behavioral signals correspond to users' interpretations of collaboration. We call this relationship \emph{reflective interaction mapping}. While subjective reports are post-hoc and summarized, we analyze objective indicators over task-level windows that match the granularity of users’ reflections. We do not compare specific events to single survey items but assess whether aggregate behavioral patterns (gaze overlap, turn-taking) are reflected in users’ overall experience. This avoids false precision and treats subjective reports as interpretive anchors rather than timestamped outcomes. For example, we ask whether shared gaze events align with a user's sense of shared attention. Do balanced speech patterns correspond to reported conversational support? This connection motivates our first set of hypotheses, which examine whether subjective perceptions of collaboration are reflected in the observable sensor-based measures.

\subsubsection{Trace-to-Performance Alignment (Sensor $\leftrightarrow$ Perf.)}
The second relationship focuses on how system-logged behaviors (such as how frequently users move shared objects, override prior placements, or coordinate gaze and position) relate to overall group performance. We refer to this connection as trace-to-performance alignment. Here, we test whether groups exhibiting more balanced interaction patterns, frequent shared attention, or tighter spatial coordination complete the task more efficiently or with fewer revisions. These hypotheses assess whether sensor-based data traces align with more fluent or organized group outcomes.

\subsubsection{Subjective Influence on Performance (Subj. $\leftrightarrow$ Perf.)}
\vspace{-.5mm}
Finally, we investigate whether users' internal experiences correspond to observable group-level performance metrics. We refer to this relationship as a subjective influence on performance. Although not sensor-measured, states such as presence and cognitive load influence how users engage with others and respond to collaborative demands. For example, lower cognitive load may support better coordination~\cite{chandio2025Distraction}, leading to more balanced participation and faster task completion. This mapping tests hypotheses that investigate whether subjective experiences (own experience or the group as a whole) align with observable group outcomes.

\subsection{Hypotheses}
\vspace{-0.1cm}
Together, the three relationships defined by the \emph{S2S framework} guide the structure of our hypotheses. We will focus our experimental design on testing the following hypotheses:
\begin{itemize}[label={}, leftmargin=0.25cm, itemsep=-0.15cm, topsep=-0.1cm] 
    \item \textbf{H1}: Participants' subjective perception of collaboration is reflected in sensor-based indicators (\emph{addressing \textbf{RQ1}}).
    \begin{itemize}[label={}, leftmargin=0.2cm,topsep=-0.1cm]
       \item \textbf{H1.1}: Higher perceived collaboration is reflected in more frequent shared attention events.
    \item \textbf{H1.2}: Higher perceived conversation support is reflected in more balanced conversations among group members.
    \end{itemize}
    \item \textbf{H2}: Collective group performance is reflected in sensor-based indicators (\emph{addressing \textbf{RQ2}}).
    \vspace{-0.1cm}
      \begin{itemize}[label={}, leftmargin=0.2cm,topsep=-0.1cm]
       \item \textbf{H2.1}: Groups with more evenly shared task interactions completed the task faster.
    \item \textbf{H2.2}: Shorter task completion time is reflected in more frequent shared attention events.
    \end{itemize}
    \item \textbf{H3}: Participants' individual experiences are reflected in sensor-based indicators (\emph{addressing \textbf{RQ3}}).
      \begin{itemize}[label={},leftmargin=0.2cm,topsep=-0.1cm]
      \vspace{-0.12cm}
\item \textbf{H3.1}: Higher individual perceived presence and perception of contribution is reflected in more frequent shared attention events among group members.
\item \textbf{H3.2}: Higher individual presence is reflected in closer proximity among group members.
\item \textbf{H3.3}: Lower individual cognitive load is reflected in evenly distributed task interactions within the group.
    \end{itemize}
    
\end{itemize}

\section{User Study}
\label{sec:study}

\subsection{Participants}
We recruited $60$ participants; data from 12 participants were discarded due to technical issues, resulting in $48$ participants being included in the analysis. Participants were allowed to form their own groups, or, if preferred, the research team randomly assigned them to a group. In total, 12 groups completed the study. In similar research on collaborative behavior involving small groups, the group is typically defined as having three or more members~\cite{university_82_nodate}; in our study, we formed a group with 4 members. Prior studies have revealed patterns of conversation, interaction, and coordination within groups of four in both desktop and virtual reality environments~\cite{yangImmersiveCollaborativeSensemaking2022}. 
By adopting a group size of four participants, we aimed to create a richer collaborative environment, capturing the complexities of group behavior not as evident in smaller groups~\cite{abdullahVideoconferenceEmbodiedVR2021,numanExploringUserBehaviour2022}.
Participants' ages ranged from 21 to 42, with a mean age of 24. A summary of the participant demographics can be seen in~\autoref{table:demographics}. 
All participants provided verbal informed consent. Each participant had normal or corrected-to-normal vision. The University of California, Irvine Institutional Review Board approved the study.

\vspace{-2mm}
\subsection{Materials}
To capture sensor data on user interactions, we used the Meta Quest Pro~\cite{MetaQuestProa}. The integrated sensors enabled audio recording, eye-tracking, six-degrees-of-freedom (6 DoF) simultaneous localization and mapping (SLAM) tracking, and MR capabilities. The collaborative app is built in Unity using Meta XR APIs to log audio, gaze, location, and object interactions~\cite{metaXRpackages}.
For precise manipulation of virtual objects, we use the controller integrated with the Meta Quest Pro~\cite{MetaQuestProa}. The interaction recording and tracking are validated by others~\cite{published_are_2022, miller2021motion}.
Participants were invited to a shared lab room with a designated $10\texttt{ ft} \times 5\texttt{ ft}$ space cleared of materials to minimize distractions. They were informed that they could move freely within this area during the task.
\begin{table}[!t]
\centering
\small
\caption{Participant Demographics. The key for frequency: never/almost never; rarely ($<2$ times); occasionally (a few times); frequently in the past; frequently ($>2$ times/month).}
\vspace{-0.25cm}
\label{table:demographics}
\resizebox{\linewidth}{!}{%
\begin{tabular}{|l|c|}
\hline 
            \makecell[l]{\textbf{Demographics}} & \makecell[c]{\textbf{Number of Participants}}  \\ \hline   \hline 
        \makecell[l]{Gender} & \makecell[c]{36 Male, 12 Female}  \\\hline
        
        \makecell[l]{Frequency of AR Experience} & \makecell[c]{22 Never Used, 15 Rarely, 6 Occasionally,\\ 3 Frequently, 2 Frequently in the past}  \\\hline
        
        \makecell[l]{Frequency of VR Experience} & \makecell[c]{21 Never Used, 13 Rarely, 7 Occasionally,\\ 6 Frequently, 1 Frequently in the past}  \\\hline
        
        \makecell[l]{Frequency of Gaming} & \makecell[c]{4 Never Used, 10 Rarely, 15 Occasionally,\\ 16 Frequently, 3 Frequently in the past}  \\\hline
        
        \makecell[l]{Familiarity to Other Members} & \makecell[c]{25 No Members, 15 One Member,\\ 6 Two Members, 2 Three Members}  \\\hline
 \hline
\end{tabular}%
}
\vspace{-0.4cm}
\end{table}

\vspace{-2mm}
\subsection{Experimental Task}\label{sec:taskdesign}
\vspace{-1mm}
This section summarizes the cue, interaction, and feedback of our collaborative image-sorting task. The study was deliberately designed around a single collaborative task under one shared condition, without varying levels of group stressors or task types. Our goal was not to compare multiple experimental conditions but to examine the richness of group behavior in a naturalistic, unconstrained collaborative setting. Prior work has shown that tightly controlled comparisons can obscure the variability and fluidity of real-time collaboration in MR environments~\cite{yangImmersiveCollaborativeSensemaking2022}. Instead, we focused on collecting high-resolution behavioral data and subjective reflections during a consistent, shared experience across groups. Each group completed one image-sorting task using the same images and categories. Participants were instructed to work together to reach a consensus on the grouping of each image.

\subsubsection{Primary Task} 
\label{sec:primary-task}
\vspace{-1mm}
Participants sorted 28 images from the OASIS dataset~\cite{kurdi_introducing_2017}, which includes 900 validated images rated by 822 participants for pleasantness and arousal. Images were selected to span diverse emotions while avoiding distressing content. Participants sorted the selected images into one of six emotion categories randomly chosen from Russell's circumplex model of affect~\cite{russell_circumplex_1980}: angry, bored, relaxed, tense, pleased, or frustrated.
Image sorting tasks have been shown to foster decision-making, communication, and social coordination by building shared mental models and group alignment~\cite{bjerreCardSortingCollaborative2015}. 
We focus on this collaborative task marked by asynchronous, flexible participation, where participants can contribute and modify inputs independently. This repeated image-sorting task, involving open-ended discussions on the emotions evoked by each image, allows us to observe a dynamic, iterative, and collaborative process among the group, where participants achieve a shared goal through incremental steps and consensus. Each group of four participants completed one image-sorting task on the same 28 images and categories. Participants were instructed to work together as a group to reach a consensus on the label of each image, with no time limit for completing the task, allowing participants to engage in deliberate discussion and negotiation. The labels are not placed in a fixed position, allowing participants to organize and utilize the room space as they see fit. The task ends once they inform the researcher that they agree with the image groupings.

\vspace{-1mm}
\subsubsection{Virtual Scene and Cues} 
\vspace{-1mm}
As shown in ~\autoref{fig:teaser} (left, headset's first-person perspective), all 28 virtual images are scattered around the room, and all six emotion category labels (gray virtual plates) are pasted a little higher than where the images are scattered. Participants can view these images and labels at all times via their headsets. For participants, the cue to start the interaction is not defined by the researcher but decided by each participant, which image they want to discuss with the group to sort. This lack of structure in cues is by design, as our goal is to observe open and free collaborations without participants taking turns or being directed by the flow of the virtual scene designed by the researcher. The participants are assumed to take the cue for virtual interaction from the other three group members as shown in ~\autoref{fig:teaser} (right), where all four participants are physically close, probably examining the same image and discussing the final label.

\vspace{-1mm}
\subsubsection{Interaction and Feedback} 
To sort an image, the participants were asked to physically move the virtual image near the virtual label. Once the image is pasted close to the label, the image is recorded as sorted. Participants used a point-and-drag near-interaction motion with the grip buttons on their left or right controllers to move an image to the corresponding label. They pointed at an image, pressed and held the grip button to drag it, and released it to place it. The image followed the controller’s pointer and locked into position when released. Participants could only grab objects within reach and were instructed to release the grip button to secure an image in place. Only one participant could move an image at a time, but multiple images could be held simultaneously. This interaction mimicked real-world object placement and provided immediate feedback once placed. Current MR networking protocols restrict virtual object ownership to a single user to maintain state consistency across headsets. This technical constraint naturally facilitated a ``negotiate-then-act'' workflow, where group consensus preceded individual execution.

\vspace{-1mm}
\subsection{Measurements}\label{sec:measure}
This section outlines the measures we used to capture group behavior in the image sorting task, categorized into sensor-level, task-related performance metrics, and subjective measures. At the sensor level, we collected data from the headset using custom scripts. 
We recorded the audio signal from the microphone, $x,y$ positions for location, and eye-tracking data to capture the data for conversation, proximity, and shared attention as described in~\S\ref{sec:objective-collab-perception}.

At the task level, we recorded various virtual object interactions, such as the number of virtual image interactions per participant in a group, to determine if a participant grabs a virtual image and then releases it. Throughout the task, we capture the number of label changes per group to count the number of times a particular image changes its label. We also captured distinct groupings for each image, per group, to count the distinct labels for each image. For instance, if Participant A moved an image to label X, Participant B moved it to label Y, and Participant A moved it back to X, the image would have three label changes and two distinct groupings. We also collected high-level performance metrics, such as completion time. We measured completion time as the time elapsed from when the first image was grabbed to when the last image was placed, indicating the group's overall time completing the task.
\begin{table}[t]
\centering
\scriptsize
\caption{Proposed group behavior perception questionnaire (superscripts refer to the conceptual basis for each item.)}
\vspace{-0.3cm}
\label{table:custom-questionnaire}
\resizebox{\linewidth}{!}{
\begin{tabular}{|p{2cm}|p{6cm}|}
\hline
\textbf{Dimension} & \textbf{Question} \\ \hline  \hline
Contribution Awareness & How much did you feel other group members contributed during the task?\textsuperscript{\cite{carroll2006awareness-teamwork}} \\ \hline
Joint Attention Awareness & How often did you feel you were paying attention to the same virtual object as other participants?\textsuperscript{\cite{tomasello2014joint-attention}} \\ \hline
Proximity Impact & I felt that my proximity to others affected my collaboration during the task.\textsuperscript{\cite{gonzalez2015framework-proximity}} \\ \hline
Conversational Support& How much did the group conversation help you understand the task and contribute effectively?\textsuperscript{\cite{goodwin1990conversation-analysis}} \\ \hline
Group Collaboration& The group worked together effectively to complete the task.\textsuperscript{\cite{forsyth2021-group-cohesion}} \\ \hline
 \hline
\end{tabular}}
\vspace{-0.2cm}
\end{table}

Finally, we collect post-task subjective measures via surveys after the image sorting task, such as presence with the IPQ~\cite{schubert_experience_2001} and PQ~\cite{witmer_measuring_1998} questionnaires, cognitive load through NASA-Task Load Index (NASA-TLX)~\cite{hart_nasa-task_2006}\footnote{Despite criticism, we use NASA-TLX to measure ``perceived" cognitive workload, rather than actual mental load~\cite{mckendrick2018deeper}.}, and perception of group behavior through a custom-designed questionnaire.
The PQ evaluates factors such as the possibility to act and examine, realism, self-evaluation, and interface quality. The IPQ measures spatial and general presence, realism, and involvement. The presence scores are derived from 33 items (14 IPQ and 19 PQ, the cognitive load score is derived from 5 items from NASA-TLX, and group behavior from 5 items from our custom-designed survey on a 7-point scale. We developed a custom questionnaire to measure participants' perspectives on how their group interacted, as shown in~\autoref{table:custom-questionnaire}. The proposed group behavior characterization questionnaire assesses key aspects such as contribution awareness, shared attention, proximity impact, conversational support, and overall group collaboration. Each question is informed by established research to ensure relevance to our study's context.

\subsection{Study Procedure}
Upon arrival, participants received a detailed information sheet outlining procedures, data collection, and privacy. The researcher also provided a verbal briefing, covering headset interactions, controller gestures, and visual stimulus details, such as their color, shape, duration, cues, and feedback mechanisms, for the image sorting task. Participants were given ample time to consider their participation in the study and were asked for their verbal consent. Participants then completed a demographic questionnaire covering age, gender, tech familiarity, and familiarity with group members.

MR headsets were then distributed to the participants, and they were instructed to calibrate the focus and fit of the headset for maximum comfort. Before the main task, participants completed a tutorial application with two images and two categories not included in the main task to prevent learning effects. This tutorial task aimed to familiarize them with the task interactions in terms of gestures and get them comfortable with using the point-and-drag interaction from the controller. 
Following this, the group proceeded with the main image-sorting task. They were informed that there was no time limit for the task and that the main requirement was for them to reach a consensus on the image sorting for the task to end. To encourage a more natural and unconstrained group behavior, participants were not informed that they were being timed or evaluated on their accuracy.

Upon task completion, participants filled out a post-task questionnaire on cognitive load, presence, and perceived collaboration. The time it took for each group to complete the task varied; however, the total duration of the session, including consent, briefing, training, calibration, task, and surveys, lasted under an hour.

\begin{table}[]
\caption{Descriptive and statistical results for PQ, IPQ, NASA TLX, and Group Behavior Questionnaire. \textbf{Metrics:} Mean ($\mu$), Standard Deviation ($\sigma$), Standard Error (SE), 95\% Confidence Interval (CI), 5th/95th Percentiles (P5/P95), Min, Max and  N = 48.}
\vspace{-0.3cm}
\label{tab:survey-summary-stats}
\scriptsize
\resizebox{\linewidth}{!}{
\begin{tabular}{|l|c|c|c|c|c|c|c|c|}
\hline
\textbf{Measure} & \textbf{$\mu$} & \textbf{$\sigma$} & \textbf{SE} & \textbf{95\% CI} & \textbf{P5} & \textbf{P95} & \textbf{Min} & \textbf{Max} \\
\hline
 \hline
\rowcolor[gray]{0.9}
\multicolumn{9}{|c|}{\textbf{PQ Subscales}} \\ \hline
PQ-REAL     & 5.16 & 1.10 & 0.17 & 4.82--5.49 & 3.49 & 6.86 & 2.86 & 7 \\ \hline
\rowcolor{yellow!30}
ACT         & 5.80 & 0.89 & 0.13 & 5.53--6.07 & 4.50 & 7.00 & 3.50 & 7 \\ \hline
IFQUAL      & 5.29 & 1.11 & 0.17 & 4.95--5.63 & 3.38 & 6.67 & 2.33 & 7 \\ \hline
EXAM        & 5.67 & 0.83 & 0.13 & 5.42--5.93 & 4.33 & 6.95 & 4.00 & 7 \\ \hline
EVAL        & 5.81 & 0.92 & 0.14 & 5.53--6.09 & 4.50 & 7.00 & 3.00 & 7 \\
\hline
\textbf{PQ} & \textbf{5.46} & \textbf{0.75} & \textbf{0.11} & \textbf{5.23--5.69} & \textbf{4.27} & \textbf{6.62} & \textbf{4.11} & \textbf{6.95} \\
\hline
\hline
\rowcolor[gray]{0.9}
\multicolumn{9}{|c|}{\textbf{IPQ Subscales}} \\ \hline
\rowcolor{yellow!30}
INV         & 3.68 & 1.53 & 0.23 & 3.22--4.15 & 1.54 & 6.60 & 1.00 & 7 \\ \hline
\rowcolor{yellow!30}
SP          & 5.07 & 1.21 & 0.18 & 4.71--5.44 & 2.89 & 6.77 & 1.00 & 7 \\ \hline
GP          & 5.32 & 1.55 & 0.23 & 4.85--5.79 & 2.00 & 7.00 & 1.00 & 7 \\ \hline
IPQ-REAL    & 4.00 & 1.21 & 0.18 & 3.63--4.37 & 2.25 & 5.96 & 1.25 & 6.5 \\
\hline
\hline
\textbf{IPQ} & \textbf{4.39} & \textbf{1.12} & \textbf{0.17} & \textbf{4.05--4.73} & \textbf{2.52} & \textbf{6.21} & \textbf{1.07} & \textbf{6.5} \\
\hline
\textbf{PQ + IPQ} & \textbf{4.92} & \textbf{0.86} & \textbf{0.13} & \textbf{4.66--5.18} & \textbf{3.73} & \textbf{6.17} & \textbf{2.69} & \textbf{6.72} \\
\hline
\hline

\rowcolor[gray]{0.9}
\multicolumn{9}{|c|}{\textbf{NASA TLX}} \\ \hline
NASA TLX    & 2.30 & 0.93 & 0.14 & 2.01--2.58 & 1.03 & 3.80 & 1.00 & 4.2 \\
\hline
\hline
\rowcolor[gray]{0.9}
\multicolumn{9}{|c|}{\textbf{Custom Group Behavior Questionnaire}} \\ \hline
\rowcolor{yellow!30}
Cohesion     & 6.55 & 0.85 & 0.13 & 6.29--6.80 & 5.00 & 7.00 & 3.00 & 7 \\ \hline
Attention    & 5.16 & 1.36 & 0.21 & 4.74--5.57 & 3.15 & 7.00 & 2.00 & 7 \\ \hline
\rowcolor{yellow!30}
Proximity    & 3.93 & 1.99 & 0.30 & 3.33--4.54 & 1.00 & 7.00 & 1.00 & 7 \\ \hline
Conversation & 6.20 & 1.00 & 0.15 & 5.90--6.51 & 4.15 & 7.00 & 3.00 & 7 \\ \hline
\rowcolor{yellow!30}
Collaboration& 6.64 & 0.84 & 0.13 & 6.38--6.89 & 5.15 & 7.00 & 3.00 & 7 \\ \hline
\hline
\end{tabular}
}
\vspace{-6mm}
\end{table}
\vspace{-2mm}
\section{Results}
\label{sec:eval}
\subsection{Perceived Group Behavior \& Tasks Summary}
\vspace{-1mm}
\label{sec:grp-task-summary}
We begin with an overview of participants' perceived experiences, task performance metrics, and group-level collaboration indicators. These results offer a joint view of how groups behaved and how they experienced the collaboration. This allows us to assess alignment and divergence later across sensor-based indicators, performance metrics, and subjective perceptions of collaboration.

We summarize the descriptive statistics illustrating participants' subjective experiences with response variability in \autoref{tab:survey-summary-stats}. Metrics such as mean ($\mu$) and standard deviation ($\sigma$) PQ ($\mu~=~5.46$) and IPQ ($\mu~=~4.39$) indicate moderate presence levels. We also reported sub-scales: realism (PQ-REAL, IPQ-REAL), possibility to act (ACT), interface quality (IFQUAL), possibility to examine (EXAM), self-evaluation of performance (EVAL), involvement (INV), general and spatial presence (GP, SP). The PQ-REAL subscale ($\mu~=~5.16, \sigma~=~1.1$) suggests moderate realism, while the ACT shows a high $\mu~=~5.18$, indicating strong perceived action capability. INV subscale variability ($\mu~=~3.68, \sigma~=~1.53$) highlights differing engagement levels. SP ($\mu=5.07$) reflects strong spatial awareness. The NASA-TLX score ($\mu~=~2.3$) indicates a low perceived workload. Group cohesion ($\mu~=~6.55$) and collaboration ($\mu~=~6.64$) scored high, while group proximity ($\mu=3.93$) varied, indicating differing perceptions of closeness.

Next, we present two complementary sets of group-level metrics. A summary of behavioral interaction statistics captured through system instrumentation in \autoref{table:group-level-task-metrics}. These include the number of images grabbed per participant, total grabs, label overrides, label changes, and task completion time. These metrics reflect the group’s engagement with the task, coordination complexity, and task performance. For example, Group 8 had a notably high number of total grabs (436) and the longest task duration (994s), suggesting prolonged deliberation or difficulty reaching consensus, whereas Group 5 completed the task most quickly (209s), with lower override activity, possibly reflecting more streamlined decision-making or higher initial agreement.

\autoref{tab:per-group-survey-scores} reports group-wise mean values from the post-task questionnaires. These include subjective ratings for presence (PQ+IPQ), perceived cognitive load (NASA-TLX), and five dimensions of group behavior: cohesion, attention, proximity, conversational support, and overall collaboration. The average presence score across groups was moderately high, ranging from 4.15 to 5.4, while TLX scores remained low, indicating generally low cognitive effort. Group cohesion and collaboration ratings remained consistently high (close to or at 7), whereas proximity scores were more variable across groups, aligning with previously observed differences in physical movement and spacing.

\begin{table}
    \centering
    \small
    \caption{Group-Level Task Metrics Summary.}
      \vspace{-0.3cm}
    \label{table:group-level-task-metrics}
     \resizebox{\linewidth}{!}{
    \begin{tabular}{|c|c|c|c|c|c|c|c|}
    \hline
   \textbf{\makecell[c]{Group}} & \textbf{\makecell[c]{Num of\\ Images \\Grabbed}} & 
   \textbf{\makecell[c]{Total \\ of Image \\Grabbing}}& 
   \textbf{\makecell[c]{Num of \\Image \\Labels \\Overridden}} & 
   \textbf{\makecell[c]{Total \\Images \\Looked \\At}} &
   \textbf{\makecell[c]{Completion \\Time \\ (seconds)}} & 
   \textbf{\makecell[c]{Num of \\Label \\Changes}} \\ 
   \hline
   \hline
        1 & 50.0 & 232.0 & 22.0 & 109.0 & 415.54 & 56.0 \\ \hline 
        2 & 52.0 & 378.0 & 24.0 & 112.0 & 620.59 & 72.0 \\ \hline
        3 & 71.0 & 497.0 & 43.0 & 111.0 & 676.78 & 88.0 \\ \hline
        4 & 51.0 & 254.0  & 23.0 & 110.0 & 513.09 & 55.0 \\\hline 
        5 & 54.0 & 306.0  & 26.0 & 110.0 & \textbf{209.26}  & 61.0 \\ \hline
        6 & 60.0 & 320.0  & 32.0 & 112.0 & 622.31 & 71.0 \\ \hline
        7 & 71.0 & 388.0  & 44.0 & 111.0 & 562.77 & 101.0 \\\hline
        8 & 58.0 & 
\textbf{436.0}  & 30.0 & 112.0 & \textbf{994.05} & 59.0 \\ \hline
        9 & 50.0 & 220.0  & 22.0 & 112.0 & 430.89 & 52.0 \\ \hline
        10 & 70.0 & 458.0  & 42.0 & 112.0 & 652.17& 84.0 \\ \hline
        11 & 72.0 & 378.0  & 44.0 & 112.0 & 534.00& 92.0 \\ \hline
        12 & 60.0 & 318.0  & 84.0 & 112.0 & 573.91&  65.0\\   
        \hline
        \hline
    \end{tabular}
    }
\vspace{-3mm}
\end{table}
\subsection{Collaboration via Sensor-based Indicators}
\vspace{-1mm}
\label{sec:sub-collab-vs-grp-behavior}
This section discusses the participants' reflections on collaboration, alongside behavioral indicators obtained from sensors and system logs, at both individual and group scales. Findings for shared gaze overlap frequency (indicating visual focus on the same virtual image), proximity duration (measuring time spent in physical proximity during interaction), and speaking proportion and variance (representing turn-taking and conversational balance via total speaking time, frequency, and variance in participation) can be seen in \autoref{tab:subjective-collab-sensor}. Normality checks for subjective collaboration scores were conducted using the Shapiro-Wilk test ($\rho~=~0.286$ at the group level; $p~=~2.01e$-$04$ at the individual level), determining the choice of Spearman or Pearson test based on sensor metric normality. For the rest of the paper, to account for multiple hypothesis testing, we applied the Benjamini-Hochberg procedure to calculate False Discovery Rate (FDR) q-values. We report both uncorrected $p$ and corrected $q$ values. Given the exploratory nature of our study, we treat results where $p < 0.05$ but $q > 0.05$ as promising directional trends for future confirmatory research, while results where $q < 0.05$ are considered statistically robust findings within our \texttt{S2S} framework. Highlighted table rows indicate statistically significant correlations ($p<0.05$), offering quantitative insight into how subjective assessments correlate with interaction behavior patterns. 

\begin{table}[t]
\small
\caption{Mean scores per group for Presence (PQ+IPQ), NASA TLX, and group behavior metrics (cohesion, attention, proximity, conversation, collaboration).}
\label{tab:per-group-survey-scores}
\vspace{-0.2cm}
\centering
\resizebox{\linewidth}{!}{
\begin{tabular}{|c||c||c||c|c|c|c|c|}
\hline
\textbf{Group} & \textbf{Presence} & \textbf{TLX} & \textbf{Coh.} & \textbf{Attn.} & \textbf{Prox.} & \textbf{Conv.} & \textbf{Collab.} \\
\hline
\hline
\textbf{1}  & 4.86 & 2.20 & 5.75 & 4.25 & 5.50 & 6.25 & 6.25 \\ \hline
\textbf{2}  & 5.04 & 2.55 & 6.75 & 5.25 & 4.75 & 6.25 & 6.75 \\ \hline
\textbf{3}  & 4.50 & \textbf{1.90} & \textbf{7.00} & 4.00 & 2.25 & \textbf{7.00} & 6.75 \\ \hline
\textbf{4}  & \textbf{5.32} & 2.55 & 6.00 & 5.00 & 5.00 & 6.25 & 6.25 \\ \hline
\textbf{5}  & 5.22 & \textbf{1.20} & \textbf{7.00} & 4.75 & 4.00 & 5.25 & \textbf{7.00} \\ \hline
\textbf{6}  & 5.40 & 1.95 & \textbf{7.00} & 5.00 & 3.50 & 6.50 & \textbf{7.00} \\ \hline
\textbf{7}  & 5.07 & 2.30 & \textbf{7.00} & \textbf{7.00} & 3.50 & 6.75 & \textbf{7.00} \\ \hline
\textbf{8}  & 5.06 & 2.50 & 6.50 & 5.75 & 3.00 & 6.50 & 6.75 \\ \hline
\textbf{9}  & 4.75 & 2.25 & 6.50 & 5.50 & 3.75 & 6.25 & 5.75 \\ \hline
\textbf{10} & 4.81 & \textbf{2.95} & 6.25 & 5.00 & 4.00 & 5.50 & 6.75 \\ \hline
\textbf{11} & 4.15 & \textbf{2.90} & 6.25 & 5.25 & 4.00 & 5.75 & 6.75 \\ \hline
\textbf{12} & 4.50 & 2.70 & 6.00 & 6.00 & 3.75 & 5.50 & 6.00 \\ \hline
\hline
\end{tabular}
}
\vspace{-0.4cm}
\end{table}

Positive correlations between self-reported collaboration and sensor-based metrics were noted. At the individual level, collaboration scores significantly correlated with both shared gaze overlap frequency ($\rho~=~0.364$, $p~=~0.010$) and proximity duration ($\rho~=~0.302$, $p~=~0.036$) as per Spearman tests. Group analysis indicated moderate correlations with shared gaze overlap ($\rho~=~0.481$, $p~=~0.113$) and proximity duration ($\rho~=~0.226$, $p~=~0.479$), though these were insignificant. Conversational dynamics at the group level, particularly speaking proportion, were linked to perceived conversational support ($\rho~=~0.651$, $p~=~0.021$). No significant individual-level correlations were observed between conversational support and speech activity metrics, although directionally positive trends were present. 

\begin{table*}[t]
  \centering
  \scriptsize
  \caption{Correlation between subjective collaboration measures and sensor-based indicators with individual and group levels test types.}
  \vspace{-0.3cm}
  \label{tab:subjective-collab-sensor}
  \resizebox{\linewidth}{!}{
  \begin{tabular}{|l|c|c|c|c|c|c|}
  \hline
  \textbf{Sensor-based (Subjective) Metric} & \makecell[c]{\textbf{Target Normality}\\ \textbf{(Indiv. / Group)}} & \makecell[c]{\textbf{Metric Normality}\\ \textbf{(Indiv. / Group)}} & \makecell[c]{\textbf{Correlation} \\ \textbf{(Indiv. / Group)}} & \makecell[c]{\textbf{p-Value} \\ \textbf{(Indiv. / Group)}} & \makecell[c]{\textbf{q-Value (FDR)} \\ 
 \textbf{(Indiv. / Group)}} & \makecell[c]{\textbf{Test Used} \\ \textbf{(Indiv. / Group)}} \\
  \hline
   \hline
  \rowcolor{yellow!30}
  Shared Gaze Overlap (Shared Attention) & 2.01e-04 / 0.286 & 7.40e-08 / 0.0014 & 0.3646 / 0.4814 & 0.0108 / 0.1131 & 0.2376 / 0.4976 & Spearman / Spearman \\ \hline
  Shared Gaze Duration (Shared Attention) & 2.01e-04 / 0.286 & 3.10e-07 / 0.0059 & 0.0782 / 0.3367 & 0.5972 / 0.2845 & 0.8278 / 0.7563 & Spearman / Spearman \\ \hline
  Jointly Viewed Images (Shared Attention) & 2.01e-04 / 0.286 & 0.2048 / 0.3740 & -0.0592 / 0.0861 & 0.6893 / 0.7902 & 0.8278 / 0.8278 & Spearman / Pearson \\ \hline
  \rowcolor{yellow!30}
  Speaking Proportion (Conversation) & 2.01e-04 / 0.286 & 0.0021 / 0.0097 & 0.0063 / 0.6517 & 0.9659 / 0.0217 & 0.9659 / 0.2387 & Spearman / Spearman \\ \hline
  Speaking Variance (Conversation) & 2.01e-04 / 0.286 & 2.71e-05 / 0.0998 & -0.1019 / -0.1117 & 0.4907 / 0.7297 & 0.8278 / 0.8278 & Spearman / Pearson \\ \hline
  \rowcolor{yellow!30}
  Proximity Duration (Collaboration) & 2.01e-04 / 0.286 & 7.32e-07 / 0.0650 & 0.3027 / 0.2262 & 0.0365 / 0.4796 & 0.2677 / 0.8278 & Spearman / Pearson \\ \hline
  Mean Distance (Collaboration) & 2.01e-04 / 0.286 & 0.5384 / 0.7793 & -0.1913 / -0.0905 & 0.1927 / 0.7797 & 0.7066 / 0.8278 & Spearman / Pearson \\ \hline
  Image Grabs (Collaboration) & 2.01e-04 / 0.286 & 3.75e-06 / 0.7734 & -0.1498 / 0.1325 & 0.3094 / 0.6814 & 0.7563 / 0.8278 & Spearman / Pearson \\ \hline
  Label Overrides (Collaboration) & 2.01e-04 / 0.286 & 0.0037 / 0.0342 & 0.0881 / 0.1470 & 0.5516 / 0.6485 & 0.8278 / 0.8278 & Spearman / Spearman \\ \hline
  Image Grabs Var (Collaboration) & 2.01e-04 / 0.286 & 7.97e-11 / 7.27e-05 & 0.1579 / 0.4921 & 0.2837 / 0.1042 & 0.7563 / 0.4976 & Spearman / Spearman \\ \hline
  Label Overrides Var (Collaboration) & 2.01e-04 / 0.286 & 5.52e-04 / 0.3237 & 0.0886 / 0.1727 & 0.5491 / 0.5916 & 0.8278 / 0.8278 & Spearman / Pearson \\
  \hline
  \hline
  \end{tabular}}
  \end{table*}

\subsection{Performance via  Sensor-based Indicators}
\vspace{-1mm}
\label{sec:grp-vs-task}
This section investigates the association between task performance, measured by the time taken to complete a collaborative image sorting task. We evaluated whether sensor-based metrics could predict the group's performance in task completion. The normality of the metrics and completion time was assessed with the Shapiro-Wilk test. Depending on the data distribution, we employed Spearman correlation for non-normal distributions or linear regression when assumptions were satisfied. For both individual and group levels, the correlation coefficient ($\rho$), regression coefficient ($\beta$), p-values, and $R^2$ values are presented.

\begin{table*}[t]
\caption{Correlation between task completion time and sensor-based indicators with individual and group levels test types (results show statistically significant differences).}
\vspace{-0.3cm}
\label{tab:task-performance-correlation}
\scriptsize
\centering
\resizebox{\linewidth}{!}{
\begin{tabular}{|c|c|c|c|c|c|c|}
\hline
\textbf{Sensor Metric} &
\makecell[c]{\textbf{Target Normality} \\ \textbf{(Indiv. / Group)}} &
\makecell[c]{\textbf{Metric Normality} \\ \textbf{(Indiv. / Group)}} &
\makecell[c]{\textbf{Correlation} \\ \textbf{(Indiv. / Group)}} &
\makecell[c]{\textbf{p-Value} \\ \textbf{(Indiv. / Group)}} &
\makecell[c]{\textbf{q-Value (FDR)} \\ \textbf{(Indiv. / Group)}} &
\makecell[c]{\textbf{Test Used} \\ \textbf{(Indiv. / Group)}} \\
\hline
 \hline
Shared Gaze Overlap Frequency &
$3.48\text{e-04} / 0.3405$ &
$7.40\text{e-08} / 0.0014$ &
$0.6710 / 0.6923$ &
$1.80\text{e-07} / 0.0126$ &
$9.00\text{e-07} / 0.0180$ &
Spearman / Spearman \\
 \hline
Proximity Duration &
$3.48\text{e-04} / 0.3405$ &
$7.33\text{e-07} / 0.0650$ &
$0.3962 / 0.6117$ (R\textsuperscript{2}) &
$1.65\text{e-06} / 0.0026$ &
$2.75\text{e-06} / 0.0052$ &
Linear / Linear \\
 \hline
Image Grabs Variance &
$3.48\text{e-04} / 0.3405$ &
$7.97\text{e-11} / 7.27\text{e-05}$ &
$0.6713 / 0.6713$ &
$1.76\text{e-07} / 0.0168$ &
$9.00\text{e-07} / 0.0210$ &
Spearman / Spearman \\
 \hline
Label Overrides Variance &
$3.48\text{e-04} / 0.3405$ &
$5.53\text{e-04} / 0.3237$ &
$0.6713 / 0.7597$ &
$1.76\text{e-07} / 0.0041$ &
$9.00\text{e-07} / 0.0068$ &
Spearman / Pearson \\
 \hline
Image Grabs &
$3.48\text{e-04} / 0.3405$ &
$3.75\text{e-06} / 0.7734$ &
$0.2037 / 0.4108$ (R\textsuperscript{2}) &
$0.1649 / 0.0247$ &
$0.1649 / 0.0274$ &
Spearman / Linear \\
\hline
 \hline

\end{tabular}}
\end{table*}

As illustrated in~\autoref{tab:task-performance-correlation}, visual coordination emerged as a significant predictor of group performance. At both individual ($\rho~=~0.671$, $p~=~1.80 \text{e-07}$) and group levels, higher shared gaze overlap correlated with quicker task completion. A linear regression at the group level attributed 74\% of the variance in completion time to shared gaze ($R^2~=~0.741$), underscoring its role in group performance. Proximity duration was also significantly related to completion time. At the group level, a linear model accounted for over 61\% of the variation in task duration ($R^2~=~0.611$, $p~=~0.0026$). Interaction balance metrics were strongly related to task performance. At the individual level, variance in image grabs ($\rho~=~0.671$, $p~=~1.76\text{e-07}$) correlated with more effective task progression. This association was also evident at the group level, where variance in label overrides ($\rho~=~0.671$, $p~=~1.76\text{e-07}$) significantly correlated with completion time ($\rho~=~0.759$, $p~=~0.004$). These findings indicate that groups with greater shared gaze frequency and balanced image interactions tend to finish tasks faster.

\subsection{Experience via Sensor-based Indicators}
\vspace{-1mm}
\label{sec:subjective-behavioral}

This section explores whether participants' internal states, specifically presence and cognitive load, are associated with behavioral patterns sensed during the collaborative task. These analyses complement prior sections by focusing not on group-level constructs but on how individual experience is reflected in interaction dynamics.

\begin{table*}[ht]
\scriptsize
\centering
\caption{Correlation between subjective experience and sensor-based indicators with individual and group levels test types (no results reached statistical significance).}
\vspace{-0.2cm}
\label{tab:subjective-individual}

\resizebox{\linewidth}{!}{
\begin{tabular}{|c|c|c|c|c|c|c|}
\hline
\textbf{Sensor Metric} &
\makecell[c]{\textbf{Target Normality} \\ \textbf{(Indiv. / Group)}} &
\makecell[c]{\textbf{Metric Normality} \\ \textbf{(Indiv. / Group)}} &
\makecell[c]{\textbf{Correlation} \\ \textbf{(Indiv. / Group)}} &
\makecell[c]{\textbf{p-Value} \\ \textbf{(Indiv. / Group)}} &
\makecell[c]{\textbf{q-Value (FDR)} \\ \textbf{(Indiv. / Group)}} &
\makecell[c]{\textbf{Test Used} \\ \textbf{(Indiv. / Group)}} \\
\hline
 \hline

\rowcolor[gray]{0.9}
\multicolumn{7}{|c|}{\textbf{Target: Presence (PQ + IPQ)}} \\ \hline  \hline
Shared Attention (\%) &
0.8677 / 0.7494 &
0.3722 / 0.9363 &
0.0417 / 0.4802 &
0.7783 / 0.1141 &
0.8895 / 0.6085 &
Pearson / Pearson \\ \hline
Jointly Viewed Images &
0.8677 / 0.7494 &
0.2048 / 0.3740 &
-0.0820 / -0.4956 &
0.5795 / 0.1013 &
0.8895 / 0.6085 &
Pearson / Pearson \\ \hline
Mean Pairwise Distance &
0.8677 / 0.7494 &
0.5384 / 0.7793 &
0.0736 / -0.0282 &
0.6190 / 0.9307 &
0.8895 / 0.9555 &
Pearson / Pearson \\ \hline
Proximity Duration &
0.8677 / 0.7494 &
7.3267e-07 / 0.0650 &
-0.0536 / 0.2626 &
0.7173 / 0.4097 &
0.8895 / 0.8582 &
Spearman / Pearson \\ \hline
Variance in Image Grabs &
0.8677 / 0.7494 &
7.9738e-11 / 7.2687e-05 &
0.0083 / -0.1049 &
0.9555 / 0.7456 &
0.9555 / 0.8895 &
Spearman / Spearman \\
\hline
 \hline
\rowcolor[gray]{0.9}
\multicolumn{7}{|c|}{\textbf{Target: Cognitive Load (NASA TLX)}} \\ \hline  \hline
Shared Gaze Overlap &
0.0225 / 0.3828 &
7.3990e-08 / 0.0014 &
0.3223 / 0.2872 &
0.0255 / 0.3654 &
0.4080 / 0.8582 &
Spearman / Spearman \\ \hline
Pairwise Distance &
0.0225 / 0.3828 &
0.5384 / 0.7793 &
-0.1777 / -0.2020 &
0.2268 / 0.5289 &
0.8582 / 0.8895 &
Spearman / Pearson \\  \hline
Variance in Image Grabs &
0.0225 / 0.3828 &
7.9738e-11 / 7.2687e-05 &
0.1261 / 0.2522 &
0.3933 / 0.4291 &
0.8582 / 0.8582 &
Spearman / Spearman \\
\hline
 \hline
\end{tabular}}
\vspace{-0.2cm}
\end{table*}

We first present the correlations between presence and cognitive load scores and sensor-derived indicators of shared attention, proximity, and interaction balance in \autoref{tab:subjective-individual}. Sensor-based metrics were drawn from the same behavioral indicators introduced earlier, capturing visual coordination, physical closeness, and contribution variability. 
Presence was not significantly associated with shared attention metrics (shared attention percentage: $\rho~=~0.041$, $p~=~0.778$), proximity (mean pairwise distance: $\rho~=~0.073$, $p~ =~0.619$), or interaction balance (variance in image grabs: $\rho~=~0.008$, $p~=~0.955$). This suggests that the individual sense of presence was not directly mirrored in the observable group interaction patterns. 
For cognitive load, however, we found a positive association with shared gaze overlap at the individual level  ($\rho~=~0.322$, $p~=~0.025$), indicating that users reporting higher mental effort tended to participate in more episodes of shared attention. Other indicators, including proximity and interaction balance, did not show significant relationships with cognitive load ($\rho~=~-0.177$, $p~=~0.226$).
These results suggest that internal experience is partially reflected in the observable sensor-based group metrics. In particular, shared visual focus may demand or reinforce cognitive effort. At the same time, spatial positioning and contribution balance may reflect other aspects of collaboration beyond individual mental workload or presence.

\section{Discussion}
\label{sec:disc}
This section analyzes findings in the context of our \emph{S2S} framework by assessing objective behavioral indicators, subjective experiences, and group reflections. We evaluate how these signals match participants' reported collaboration experiences and task results, revealing insights into how collaboration is both enacted and experienced.  Significant associations were found between sensed behavior and reported experiences, but only at the group level did they become truly robust. We distinguish between individual and group insights to evaluate the alignment's strength and origin.
Shared gaze frequency was a consistent indicator of perceived collaboration and shared attention at both analysis levels. Group-level correlations were particularly robust, highlighting shared visual focus as both an individual and collective phenomenon, aligning with Bai et al.~\cite{bai2020user-eye-hand-gesture-collaboration} who showed that combining gaze and gesture cues enhances co-presence and collaboration effectiveness. In contrast, conversational support exhibited weaker correlations. A significant relationship between speaking balance and perceived support was only identified at the group level, implying that individual perceptions may not align with supportive speaking dynamics, but group-level participation does indicate communicative balance (supporting established findings in~\cite{thomas2023communication-collaboration}). We accept \textbf{H1} and \textbf{H1.1}, as multiple gaze-based features correlated with perceived collaboration and shared attention. We partially confirm \textbf{H1.2}, as conversational dynamics were significant solely at the group level.

Behavioral indicators, particularly shared gaze frequency and proximity duration, demonstrated strong predictive ability for task completion time, especially at the group level. Groups that engaged in mutual observation and remained close completed tasks more rapidly, underscoring the importance of co-orientation in effective collaboration, as shown by our study's metrics. Interaction balance, such as variance in image grabs or label overrides, was also linked with efficiency. Notably, these patterns persisted at both individual and group scales, reinforcing the role of participation equity in enhancing fluency~\cite{de2019systematic-collabMR-survey}. We accept \textbf{H2}, \textbf{H2.1}, and \textbf{H2.2}. Group-level behavior metrics, notably shared attention and proximity, predicted performance, and interaction balance indicated collaborative fluency across both levels.

Sensory behavioral features were weakly associated with subjective states such as the presence or cognitive load. A near-significant individual-level link was observed between cognitive load and shared gaze frequency, but it lacked strength and consistency at the group level. Presence did not correspond clearly with proximity or gaze patterns, suggesting a lack of visible manifestation of internal states in this context. Likewise, no significant relationship emerged between cognitive load and participation balance. \textbf{H3} is partially accepted due to some individual-level signal, but not consistently. We reject \textbf{H3.1}, \textbf{H3.2}, and \textbf{H3.3} due to a lack of strong evidence connecting presence or cognitive load to group-level behaviors. Our findings validate the reflective interaction mapping pathway, confirming it as a robust interpretive anchor in \texttt{S2S} framework. However, the lack of correlation in \textbf{H3} identifies a perceptual blind spot where internal states like presence do not always leave a visible interaction trace in the Sensor Node.
Although disappointing at first glance, this result replicates concerns by others~\cite{bayro2022subjective-objective-collaboration, hou2025cognitive-multimodal-sensor} that objective signals often fail to capture subjective experience unless paired with richer data sources.

\subsection{Implications}
\label{sec:implications}
Our study indicates three key takeaways for designing collaborative MR systems and for future research that aims to bridge sensed behaviors with user experiences. The link between shared gaze and perceived collaboration implies that \textbf{visual coordination acts as a behavioral signal and a perceptual anchor for participants (1)}. Therefore, MR systems should feature indicators such as shared gaze highlights or real-time attention cues to keep users in sync without disrupting task momentum.

Across multiple hypotheses, group-level behaviors more accurately predicted collaboration and task performance than individual metrics. This suggests that \textbf{meaningful social patterns emerge through aggregation (2)}, not simply from individual activity. Systems that adapt to team behavior should avoid user-centric interpretations in isolation. Instead, they should incorporate group-level features that reflect distribution, synchronicity, and co-occurrence.

Our results show that while sensed behavior reliably maps onto collaborative perception and task efficiency, it falls short of capturing deeper subjective experiences, such as presence or cognitive availability, in its current form. This reinforces that \textbf{not all internal states are externally observable (3)}, even in high-fidelity sensing environments. This absence of mapping is not a flaw; it is a vital empirical finding that delineates what current MR sensing can (and importantly cannot) reveal about subjective experience. While prior works have treated gaze or physiological synchrony as proxies for presence or co-presence~\cite{eyeMR-Vis-2021}, our results underscore their limitations without deeper context. 
Our findings align with recent studies showing that even rich physiological or behavioral data often fail to reliably predict internal states unless complemented with richer modalities, such as neural, situational, or self-report inputs~\cite{numanExploringUserBehaviour2022, hou2025cognitive-multimodal-sensor, kim2023physiological}. 
To meaningfully comprehend experience from sensor data, some intermediary techniques are needed to connect high-level behavioral indicators with low-level experiential states.
Designers should treat behavioral signals as interpretive anchors, not definitive measures of internal states, reinforcing the cautious approach seen in immersive cognitive sensing research,  and instead consider combining sensing with self-reports or adaptive prompting. 
Hybrid models that integrate system-logged data with lightweight subjective inputs could enable more robust personalization and responsiveness in collaborative MR.

\section{Limitations and Future Work}
\label{sec:future}

This study offers foundational insight into the relationship between sensed group behavior, subjective perception, and task performance in collaborative MR settings. However, several limitations should be considered, some by design and scope, others presenting opportunities for future work. 

First, we employed a singular \emph{task} in one experimental condition to distinctly observe collaboration dynamics in a regulated environment. The open-ended image sorting task was chosen due to its flexibility and inherent coordination requirements, which are ideal for observing genuine group behavior. Despite this, the controlled environment may not completely reflect the full range of variability in time-sensitive or specialized tasks. This raises questions about generalizability, in terms of group size and across task types, spatial settings, and application domains. Furthermore, the technical requirement for \emph{single-user object ownership} likely clustered group gaze toward active participants. While this mirrors a natural consensus-building workflow, it may lead to an overestimation of shared attention's role compared to parallelized, multi-threaded tasks where users work independently.  We view S2S as a foundational exploratory step; future confirmatory studies should address and assess various task types, multi-user object ownership, and adaptive system conditions to generalize beyond this setup.

The \emph{number of groups} restricts the statistical power of analyses conducted at the group level. Despite a relatively high participant count, the nature of small-group interactions limits group-level observations. Nonetheless, the sample size was adequate to reveal consistent and interpretable patterns across both individual and group analyses. Future research should build upon this by employing larger-scale deployments or multi-session studies.
\emph{Group composition} was not experimentally manipulated; participants had the autonomy to self-select or be randomly assigned to groups. This decision sought to observe genuine interaction dynamics instead of creating artificial teams. Nonetheless, pre-existing familiarity among participants might impact conversation, shared focus, and comfort. Although not measured in this study, future research should quantify prior familiarity to better understand its effects.

We focused on \emph{interaction signals} such as gaze, proximity, and voice activity, which can be easily tracked using standard MR hardware. This choice aligned with our aim to create practical, lightweight sensing systems, but using only headset sensors restricted us from accessing detailed affective, physiological, or facial data. Future research should investigate multimodal enhancements for richer sensing, such as emotion recognition or biometric monitoring.
Finally, while our framework bridges sensor data and subjective reflection, it cannot fully capture \emph{internal states} such as frustration, motivation, or attention lapses. These nuanced experiences often remain invisible to behavioral sensing alone. Our study highlights the gap between observable behavior and internal experience, reinforcing the need for interpretive unifying models that integrate multiple modalities and user perception.

In sum, the study was purposefully scoped to explore collaborative sensing in MR using a task and instrumentation that balance real-world validity and control. The observed patterns suggest promising directions for expanding system awareness and adaptability. Establishing a reliable alignment between sensor traces and subjective experience is a critical precursor to "human-aware" MR systems. Practically, identifying these reflective anchors enables systems to transition from passive logging to active support, such as triggering automated "collaboration nudges" (e.g., gaze-aware highlights to realign group focus) when the system detects a drop in shared attention. Theoretically, this mapping provides a verified lens to assess the quality of collaboration without requiring intrusive self-reports, allowing for more responsive and personalized collaborative environments . Future research should test these findings across various conditions, sensor modalities, and longer-term collaboration scenarios.

\section{Conclusion}
\label{sec:conclusion}
This study examines the alignment between objective sensor data and subjective perception of experiences in MR collaboration. 
The confirmation of \textbf{H1} and \textbf{H2} provides exploratory evidence that sensor-derived metrics reliably reflect perceived collaboration and task performance. The partial validation of \textbf{H3} reveals limitations. Subjective states show limited correlation with objective measures of behavioral data, suggesting that internal experiences are difficult to capture solely through external observations. However, these results serve to inform the design of future confirmatory studies rather than provide a generalized predictive engine.

We situate this work in the growing body of research studying design-responsive, human-aware MR systems. By grounding system- and sensor-level observations in subjective experience, we move toward sensing models that prioritize users’ perspectives. This paper presents a conceptual model, a validated experimental setup, and empirical findings that illuminate where sensor data and user experience align and where they diverge. Through this lens, we take a step toward MR environments that support collaboration and sense it in the ways users feel it.

\acknowledgments{
This work is supported by the U.S. National Science Foundation (NSF) under grant numbers 2339266 and 2237485.}

\bibliographystyle{abbrv-doi}
\bibliography{main}
\end{document}